\documentclass[manuscript]{emulateapj}

\def\msol{\,{\rm M}_\odot}              
\def\msink{\,{\rm M}_{\rm sink}} 
\def\mcore{\,{\rm M}_{\rm core}} 
\def\msup{\,{\rm M}_{sup}} 
\def\minf{\,{\rm M}_{inf}} 
\def\tff{\,{\rm t}_{\rm dyn}}

\def\ga{\,\hbox{\hbox{$ > $}\kern -0.8em \lower 1.0ex\hbox{$\sim$}}\,}
\def\la{\,\hbox{\hbox{$ < $}\kern -0.8em \lower 1.0ex\hbox{$\sim$}}\,}

\begin{document}

\title{Star formation: statistical measure of the correlation between the prestellar core mass function and the stellar initial mass function}

\author{Gilles Chabrier}
\affil{Ecole Normale Sup\'erieure de Lyon,
CRAL (UMR CNRS 5574)\\ 69364 Lyon Cedex 07,  France.\\
School of Physics, University of Exeter, Exeter, UK EX4 4QL}

\and

\author{Patrick Hennebelle}
\affil{Laboratoire de radioastronomie, UMR CNRS 8112,\\ Ecole normale sup\'erieure et Observatoire de Paris,\\
24 rue Lhomond, 75231 Paris cedex 05, France }

\date{}


\begin{abstract}
We present a simple statistical analysis of recent numerical simulations exploring the correlation between the core mass function obtained from the fragmentation of a molecular cloud and the stellar mass function which forms from these collapsing cores. Our analysis shows that the distributions of bound cores and sink particles obtained in the simulations are consistent with the sinks being formed predominantly from their parent core mass reservoir, with a statistical dispersion of the order of one third of the core mass. Such a characteristic dispersion suggests that the stellar initial mass function is relatively tightly correlated to the parent core mass function, leading to two similar distributions, as observed. This in turn argues in favor of the IMF being essentially determined at the early stages of core formation and being only weakly affected by the various environmental factors beyond the initial core mass reservoir, at least in the mass range explored in the present study.
Accordingly, the final IMF of a star forming region should be determined reasonably accurately, statistically speaking, from the initial core mass function, provided some uniform efficiency factor. The calculations also show that these statistical fluctuations, due e.g. to variations among the core properties, broaden the low-mass tail of the IMF compared with the parent CMF, providing an explanation for the fact that this latter appears to underestimate the number of "pre brown dwarf" cores compared with the observationally-derived brown dwarf IMF.
\end{abstract}

\keywords{stars: formation --- stars: luminosity function, mass function --- ISM: clouds }

\section{Introduction}
Understanding star formation and the origin of the stellar initial mass function (IMF) remains a major challenge in astrophysics. Various observations have suggested a strong similarity between the IMF and the mass function of gravitationally bound structures in molecular clouds, identified as the prestellar {\it core} mass function (CMF), the first one being shifted downwards compared to the second one by a nearly
mass-independent factor of about 2-3 (Motte et al. 1998, Testi \& Sargent 1998, Johnstone et al. 2000, 2001, Andr\'e et al. 2007, Alv\'es et al. 2007, Nutter \& Ward-Thompson 2007, Simpson et al. 2008, Enoch et al. 2008, Andr\'e et al. 2009, 2010). These observations suggest that
the IMF is essentially determined by the properties of the turbulent self-gravitating gas in the parent molecular cloud, which leads to the core formation. Then, magnetically-driven outflows are probably responsible for the subsequent mass-loss between the core mass and the stellar mass,
yielding the aforementioned $\sim 30$-$50\%$ efficiency factor characteristic of the CMF to IMF evolution (Matzner \& McKee 2000). An analytical theory for the formation and the mass distribution of unbound overdense "clumps" and gravitationally bound "cores" directly inherited from the global physical properties of the molecular cloud
has recently been formalized by Hennebelle \& Chabrier (2008, HC08; 2009, HC09) and has received some support from numerical simulations of compressible turbulence aimed at exploring this issue (Schmidt et al. 2010). 
 
 Alternatively, some authors (Bate \& Bonnell 2005 and references therein) have suggested that the CMF is essentially determined by the various environmental conditions in the star forming region (nearby massive stars, competitive accretion between prestellar cores, dynamical interactions,...), i.e. by the various processes converting gas into stars. Accordingly, these authors argue that there is no correlation between the CMF and the IMF, as any possible link between these two distributions will inevitably be wiped out by these various environmental factors. Another notable difference between these two scenarios of star formation is the reason for the universal behaviour of the IMF and the nearly invariance of the location of the peak of the CMF/IMF in various star forming regions.
In the first scenario, this universal property arises from the universal
 behaviour of the turbulent spectrum, which tends to form clouds with similar (Larson-like) properties\footnote{Remember that in the Hennebelle-Chabrier theory, the slope of the IMF and the Larson coefficients are directly related to the turbulence power spectrum index (HC08).}, while the peak invariance arises from the similar but opposite dependence of the Jeans mass and Mach number upon the
 cloud's size (HC08, HC09). In contrast, in the second scenario, the IMF universality stems from competitive accretion and dynamical interactions,
 which wipe out the initial conditions due to the cloud's properties. 
  
Finding out which one, if any, of these scenarios is the dominant one, and thus whether or not there is a direct correlation between the CMF obtained from the fragmentation of a molecular cloud and the IMF which forms from these bound cores, represents a major issue to understand the very nature of star formation.
Recently, Smith, Clark and Bonnell (2009, hereafter S09) have conducted dedicated numerical simulations in order to explore this issue. 
The original core masses in the simulations are identified from their gravitational potential whereas the final "stellar" masses are identified by sink particles. Because of the dispersion in the final sink mass distribution, these authors argue that there is a poor correlation between the IMF and the CMF, and use this argument to invoke environmental conditions, i.e. accretion from the surrounding medium, for being a dominant factor in the determination
of the final stellar IMF. This conclusion bears important consequences to determine whether or not
the final IMF of a star forming region can be predicted accurately from the observed core mass distribution.
Using core mass distributions similar to the ones obtained by S09 and simple statistical calculations, we determine the statistical correlation between the CMF and the sink MF obtained in these simulations, in terms of a mass variance characteristic of the width of the CMF-IMF dispersion.

\section{Calculations}
\label{theory}

The simulations of S09 identify two types of structures, as the outcome of the fragmentation of a molecular cloud. The initial
overdense structures (denominated "p-cores" by the authors) are identified as peaks in the gravitational potential compared to the surrounding background. Note that some of these "p-cores" can have enough internal energy to prevent collapse and are thus unbound transient structures. Tracing the core binding energy throughout the structure lifetimes, S09 identify the {\it bound} cores as the structures with positive binding energy,
supposed to represent the gravitationally bound prestellar cores observed in mm-surveys. About 300 bound cores are found in the simulations, out of 573 initial p-cores. Eventually, these bound cores will collapse or fragment into smaller structures, identified in the simulations 
as sink particles, which are supposed to represent stars. 

We start with a random distribution of about 300 bound cores within a mass range $[\minf,\msup]=[0.2,2]\msol$, similar to the number and mass range of bound cores in the simulations of S09\footnote{The 300 core calculations allow a direct comparison with the S09 results, but we have verified the robustness of our conclusions by conducting calculations with $10^5$ cores.}. In order to be consistent with the numerical simulations, the core masses are drawn randomly according to a probability law ${\cal P}(m)=\int_{\minf}^m p_1(x)\,dx$, where the probability density $p_1(x)$ is given by the Salpeter mass function over the entire mass range, $p_1(x)\propto x^{-2.35}$. Our core population is thus consistent with the one identified in S09 (see their Fig. 5).

\noindent In order to measure, statistically speaking, the degree of correlation (or lack of) between the initial CMF and the final IMF, we consider
the probability to form a sink of mass $\msink$ from a mass reservoir of mean value $\mu=\epsilon\times \mcore$ to be given by a normal (gaussian) law, of probability density $p_2(x)=\exp[-(x-\mu)^2/2\sigma^2]$. Given the large statistical uncertainties entering the core identification and properties (contributions of the different modes of turbulence leading to the velocity dispersion, density distribution within the core, shape of the gravitational potential,...),
it seems acceptable, in the absence of a more accurate study of systematic effects, to invoke such a normal density probability on the basis of the central limit theorem (e.g. Adams \& Fatuzzo 1996). According to the above probability law, the probability for a sink mass $\msink$ to be drawn from a reservoir of  mean mass $\epsilon\times \mcore$, with some characteristic mass variance $\sigma^2$, is thus given by:.

\begin{eqnarray}
\msink = y\cdot \sigma + \epsilon (t)\times \mcore, 
\label{normal}
\end{eqnarray}
where the core mass $\mcore$ is sampled from the aforementioned Salpeter distribution, and $y$ is a random variable of mean $0$ and variance $1$.
The factor $\epsilon (t)$
illustrates the mass fraction
accreted from the core mass reservoir onto the sink within a given time $t$.
Note that the simulations of S09 do not include mass loss events such
as outflows in the core-to-sink conversion, so that the sink masses correspond to the maximum amount of accreted mass within a time $t$.
The variance $\sigma^2$ illustrates the fluctuations due essentially to variations in the intrinsic core properties or core-to-sink evolution (see above). 
In the absence of a complete theory or numerical simulations able to accurately determine such a variance, we can infer it by comparing the sink distributions obtained from eqn.(\ref{normal}) with the results of S09, for various values of $\sigma$.
Recent simulations (Dib et al. 2007) show that surface and volume energies contribute similarly to the virial balance of gravitationally bound prestellar cores. Since the mass of these collapsing cores is indeed determined by the virial condition (Tilley \& Pudritz 2004, Dib et al. 2007, HC08), it thus seems reasonable to take the variance due for instance to the fluctuations in the shape of the cores to be {\it of the order of} half the core mass. As shown below, a value $\sigma \simeq \mcore/3$ yields the best agreement with the S09 results. We will thus
pick this value as the fiducial characteristic mass variance of eqn.(\ref{normal}).

\begin{figure}[p]
\center{\includegraphics[angle=0,width=6in]{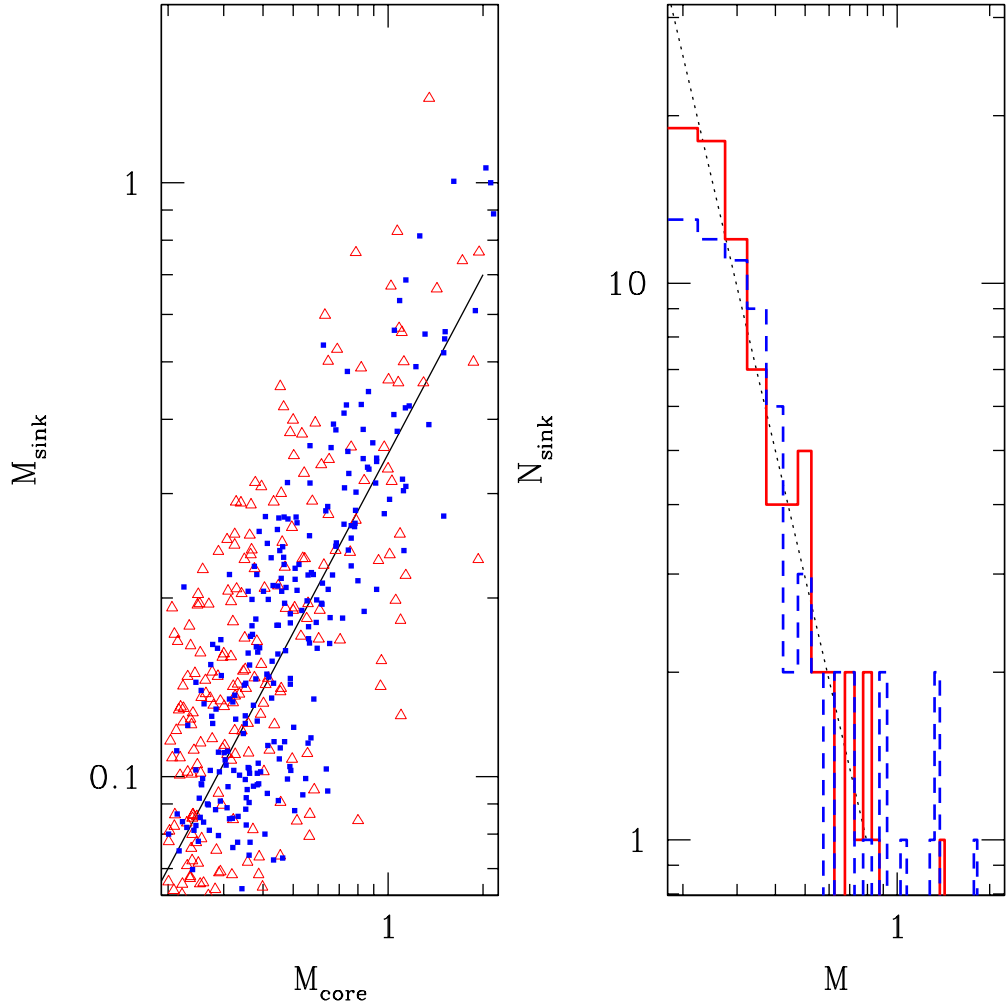}} 
\caption{Left panel: sink mass distribution obtained from the core mass distribution according to eqn.(\ref{normal}) for $\epsilon=0.3$ and $\sigma = \mcore/3$ (red triangles). The small blue crosses represent the sinks identified in the simulations of Smith et al. (2009), kindly provided by Rowan Smith, for $t=t_{dyn}$. The solid line corresponds to $\msink=\epsilon\times\mcore$. Right panel: the two respective mass distributions (solid line: present; dashed line: Smith et al. (2009). Dotted line: Salpeter mass function.}
\label{dist1}
\end{figure}

The comparison between the sink mass distributions obtained from eqn.(\ref{normal}), with $\sigma = \mcore/3$,  and S09 (their Fig. 10) at $t=\tff$ is portrayed in Figure \ref{dist1}. 
The solid line corresponds to a "perfect" (zero variance) correspondence between the core mass and the sink mass, for a global uniform efficiency factor $\epsilon=0.3$. The figure illustrates the similarity between the two distributions. This is quantified on the right panel of the
figure, which portrays the two sink mass distributions: both distributions agree within less than
one Poissonian fluctuation over the mass range presently probed by the simulations.

These results show that the sink masses obtained from the numerical simulations of a collapsing
molecular cloud are consistent with these masses being predominantly determined by the initial core mass reservoir, with some inherent statistical dispersion characterized by a standard deviation of the order of 1/3 of the core mass.
The time evolution of this deviation can be inferred from a comparison of Fig. 10 of S09, which displays the sink mass distribution for various dynamical times, and the sink mass distribution obtained from eqn.(\ref{normal}) for various values of the coefficient $\epsilon$, which kind of mimics a time sequence evolution in the CMF-to-IMF conversion. Such a comparison is portrayed in Fig. \ref{dist2}, where the value of $\sigma$, which determines the width of the distribution, is determined by looking for the best agreement between the two distributions. As seen in the Figure, only after about 3 dynamical times in the S09 simulations, does the standard deviation start to increase from the fiducial value $\mcore/3$, although even after 5 dynamical times, the deviation remains of the order of about 0.5-0.6 core mass. However,
as discussed in the conclusion, exploring the CMF-to-IMF process after several dynamical times is probably unrealistic in reality as magnetically driven outflows will halt the accretion, and thus the star formation long before.


The sink mass function obtained according to eqn.(\ref{normal}) is portrayed in Fig. \ref{figMF}, for $ \epsilon=0.3$ and $\epsilon=1$, as well as the CMF sampled
from the Salpeter-like probability law. It is clear that the sink mass function closely resembles the CMF and recovers a Salpeter IMF. This was indeed anticipated from eqn.(\ref{normal}). Since, however, this sink mass distribution agrees well with the one obtained in S09 simulations, arising from the star-forming gas cores produced by molecular cloud fragmentation, it is tempting to suggest that
the observed resemblance
between a prestellar CMF and the resulting stellar IMF indeed arises from the strong correlation, statistically speaking, between the two distributions, characterized by a mass variance $\sigma^2\simeq(\mcore /3)^2$, and a $\sim 30\%$ or so uniform efficiency factor, as suggested by
observations and theoretical calculations (see the Introduction).


\subsection{Contribution from the core mass ranges}

Following S09 (see their Fig. 11 and 12), we explore the contribution to the final sink mass distribution arising from various core mass ranges. Figure \ref{distMF} (top) portrays the
CMF obtained from the Salpeter-type probability law, within the same characteristic mass range as the simulations, with various symbols and colors denoting different mass domains. Figure \ref{distMF} (bottom) illustrates the sink MF obtained with the gaussian probability law given by eqn.(\ref{normal}), for $\epsilon=1$ and $\sigma=\mcore/3$, corresponding to  2 $\tff$ in S09 simulations (see Fig. 2). The symbols/colors of the sinks are the same as their respective parent cores.
The colors in the sink MF are indeed well mixed, as found in the simulations of S09. Therefore, correlation between the CMF and the IMF does not imply that the IMF reflects the very same distinct color/symbol domains as the CMF. Statistical fluctuations within the core properties, characterized by the mass variance, lead to the color/symbol mixing of the sink IMF. Indeed, a given core mass leads eventually predominantly to a sink mass of similar value (for $\epsilon=1$) but with some dispersion, producing a "spread" of the initial core mass domain over a larger final sink mass one, as illustrated in the figure. 


\subsection{Affecting the width of the IMF}

It has been suggested in HC08 and HC09 that the theoretical CMF might underestimate the number of (pre)brown dwarfs compared with the observationnally-derived Chabrier (2003) system IMF. Although this statement should be taken with caution, as the space density of field or even young cluster brown dwarfs remains very uncertain, we show below that the aforementioned statistical fluctuations due to variations among the core properties naturally lead to a broader distribution in the low-mass part of the IMF and thus provide a natural explanation for the (supposed) lack of low-mass brown dwarfs predicted by the HC CMF. This is
illustrated in Fig.\ref{figMFbroad}. The CMF is drawn (with $10^5$ particles) from a probability density given by a lognormal form slightly narrower than the Chabrier system IMF ($\sigma=0.4$ instead of 0.57), for the same mean mass, in order to roughly mimic the low-mass part of the theoretical Hennebelle-Chabrier CMF (see e.g. Fig.5 of HC08 or Fig. 8 and 9 of HC09). The sink mass distribution is obtained according to eqn.(\ref{normal}) for $\epsilon=1.0$ and $\sigma=0.5\times\mcore$ (blue dots) or $\sigma=0.8\times\mcore$ (red crosses).
These sink mass spectra are broader than the original CMF and recover reasonably well the Chabrier IMF.


\section{Conclusion}

\begin{figure}[p]
\center{\includegraphics[angle=0,width=6in]{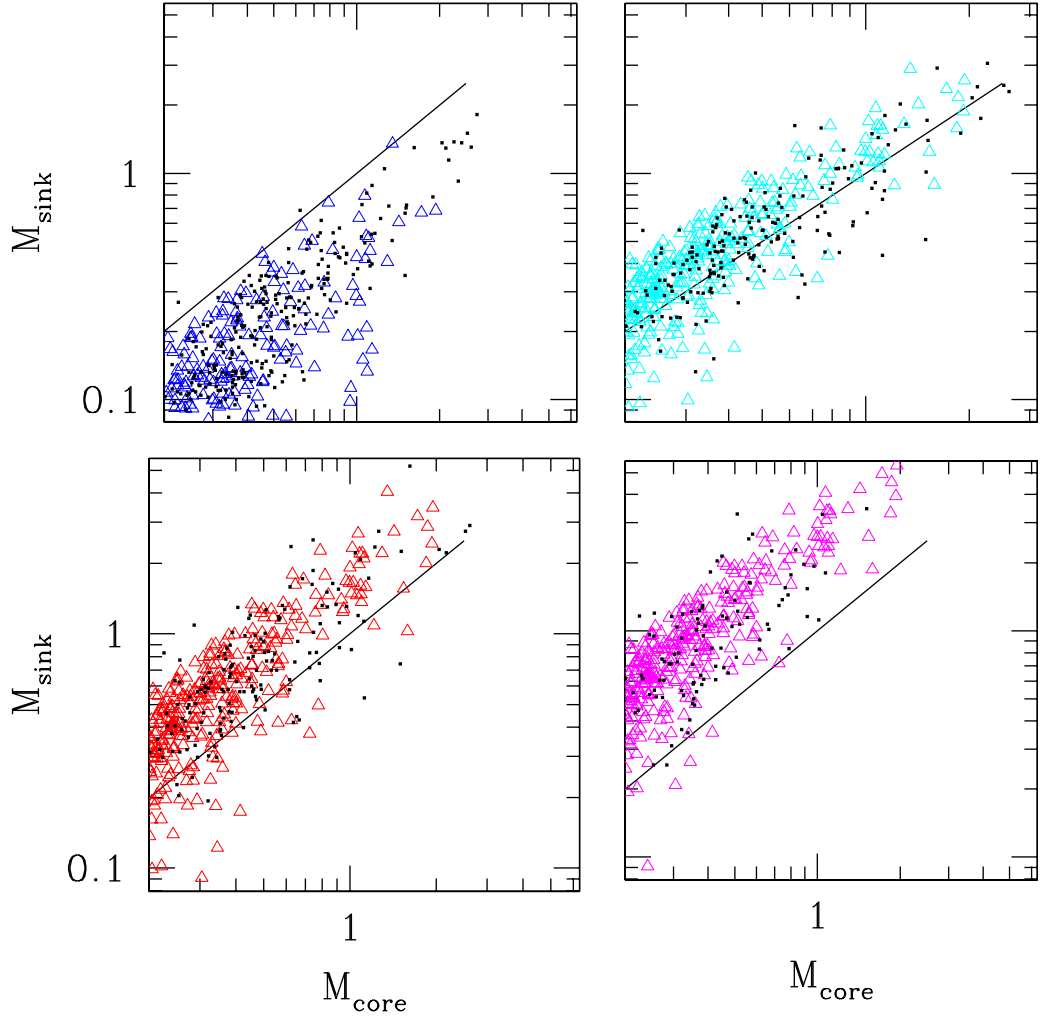}}  
\caption{Same as Fig. \ref{dist1} for various values of $\epsilon$, namely $\epsilon=0.3,1, 1.5$ and 2 from top left to bottom right, and $\sigma = 0.3\times\mcore,\,0.3\times\mcore,\,0.5\times\mcore$ and $0.6\times\mcore$,  from top left to bottom right, to be compared with the results of 1, 2, 3 and 5 $\times \tff$ in Smith et al. (2009, Fig. 10), respectively, illustrated by the small dots. The solid line corresponds to $\msink=\mcore$.}
\label{dist2}
\end{figure}

\begin{figure}[p]
\center{\includegraphics[angle=0,width=6in]{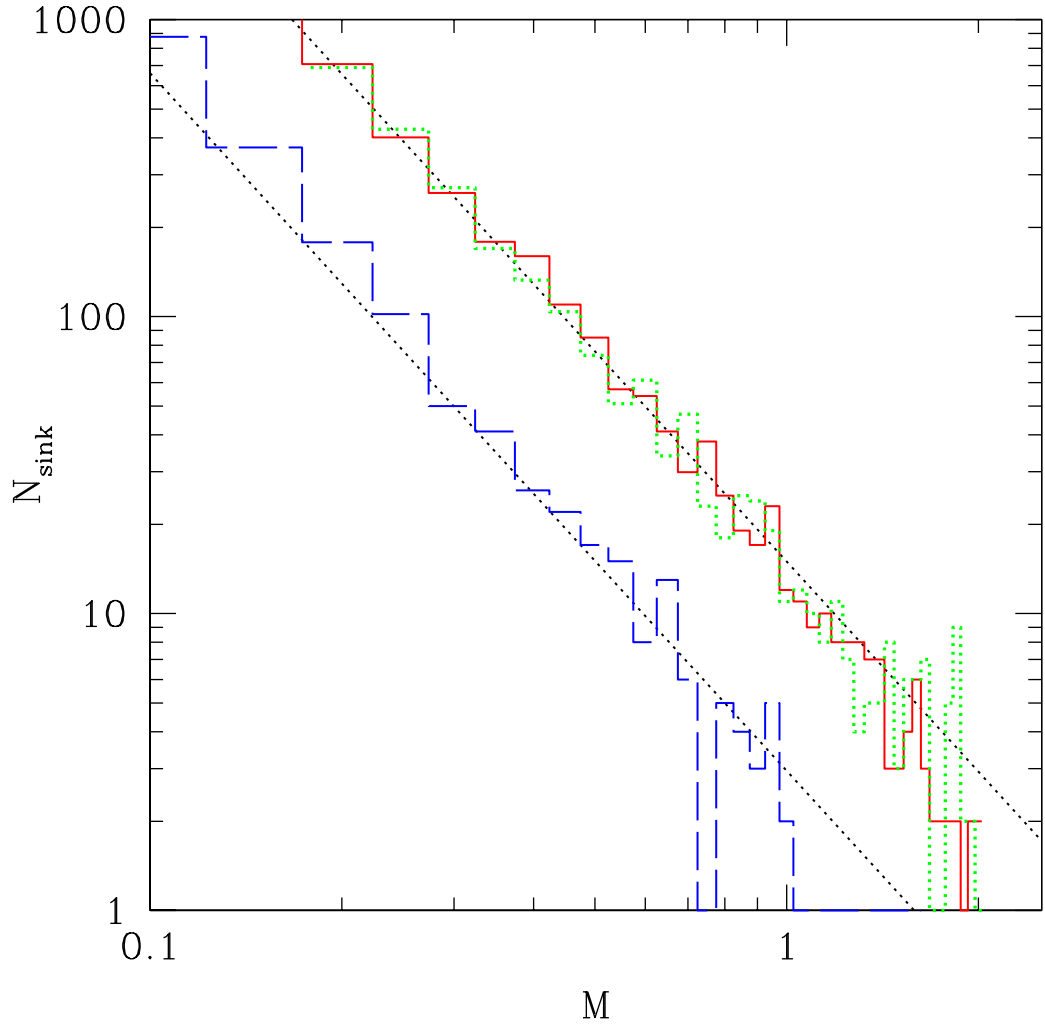}}  
\caption{Dotted (green) line: core mass distribution obtained from a Salpeter-like probability law. Solid (red) line: sink mass distribution
obtained according to eqn.(\ref{normal}) for $\epsilon=1.0$; long-dash (blue) line: same for $\epsilon=0.3$. Diagonal dotted line: Salpeter IMF. For better statistics, the calculations have been conducted with a sample of 5000 core masses.}
\label{figMF}
\end{figure}

\begin{figure}[p]
\center{\includegraphics[angle=0,height=7in,width=6in]{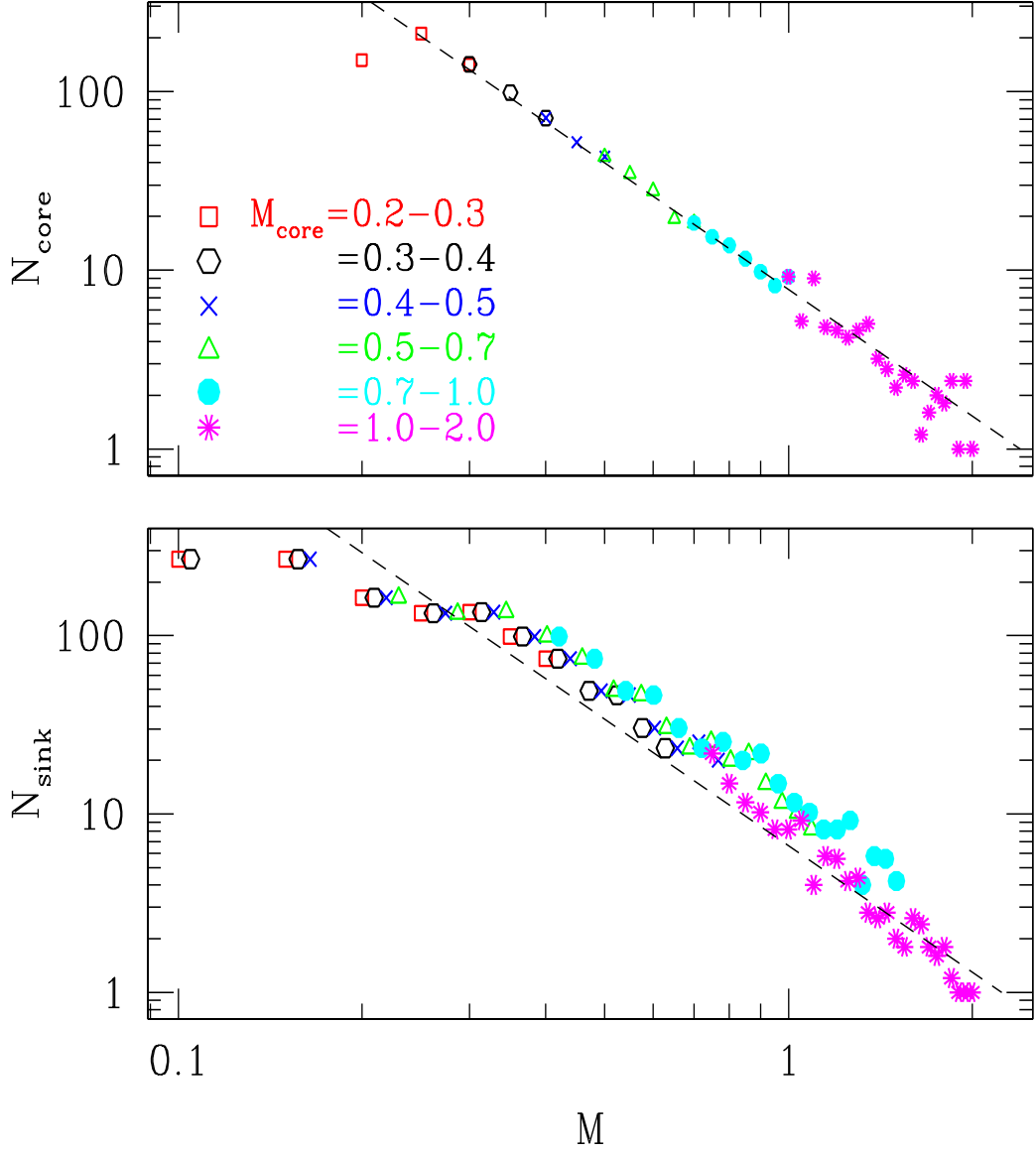}}  
\caption{Top panel: mass distribution for the cores sampled from a probability distribution given by a Salpeter MF, over a 0.2-2.0 $\msol$ mass range. The different symbols/colors correspond to various mass ranges: squares (red): 0.2-0.3 $\msol$;  empty circles (black): 0.3-0.4 $\msol$; crosses (blue): 0.4-0.5 $\msol$; triangles (green): 0.5-0.7 $\msol$; full circles (cyan): 0.7-1.0 $\msol$; stars (magenta): 1.0-2.0 $\msol$. Bottom panel: sink mass distribution obtained from the above core MF, according to the probability law given by eqn.(\ref{normal}), with $\sigma=\mcore/3$. The colors/symbols correspond to the aforementioned parent core masses. For a given mass, the various symbols/colors illustrate the different parent core mass domains contributing to the total number N$_{sink}$;
for a better identification, each contributing mass domain has been slightly shifted rightwards, yielding a slight shift of the sink MF. Short-dashed line: Salpeter mass function.}
\label{distMF}
\end{figure}

\begin{figure}[p]
\center{\includegraphics[angle=0,width=6in]{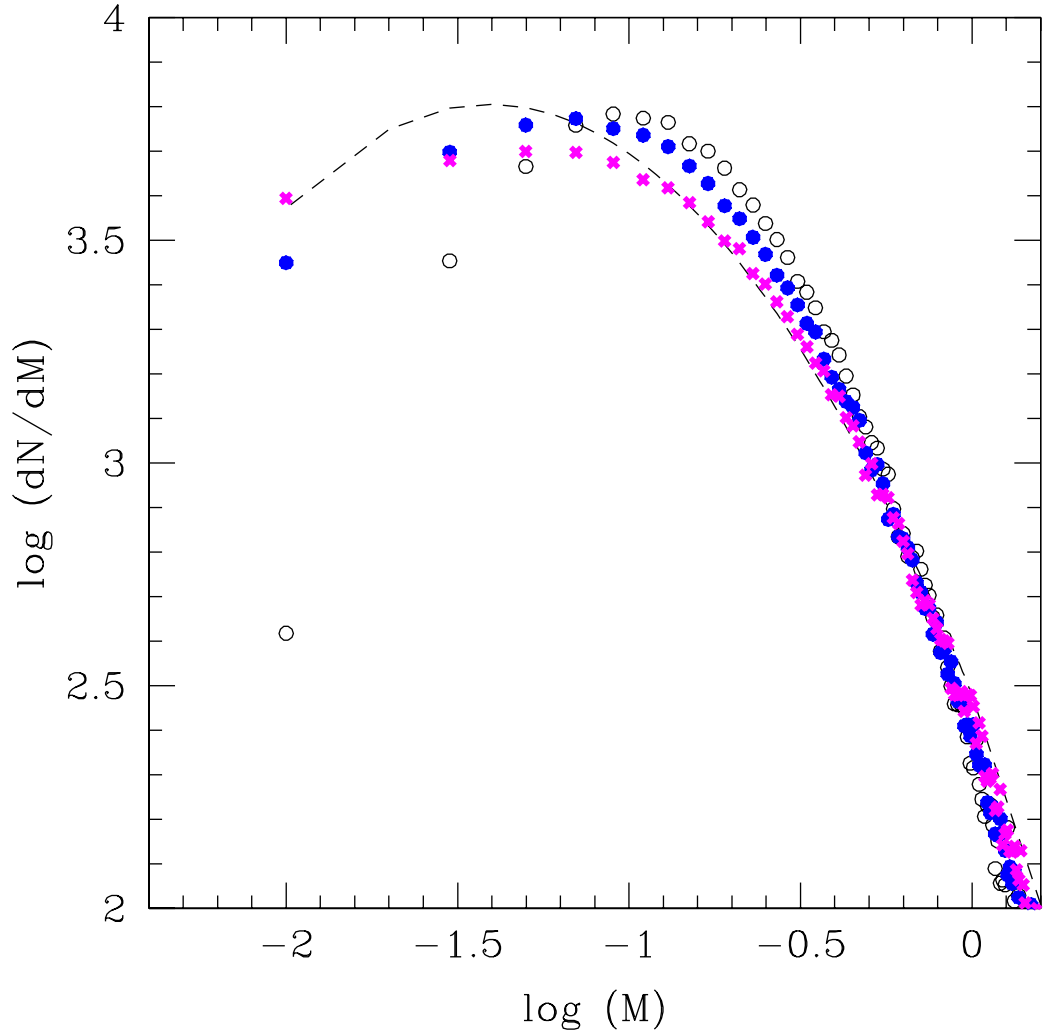}} 
\caption{Empty circle: core mass spectrum drawn from a lognormal probability distribution with a mean $\langle \log \,0.22 \rangle$ (Chabrier 2003) and a standard deviation $\sigma=0.4$; Dots and crosses: sink mass spectrum obtained from this core mass spectrum according to eqn.(\ref{normal}) for $\epsilon=1.0$ and $\sigma=0.5\times\mcore$ (blue dots) and $\sigma=0.8\times\mcore$ (red crosses); Dash-line: Chabrier (2003) system IMF.}
\label{figMFbroad}
\end{figure}

In this Letter, we have examined the degree of correlation between the prestellar CMF and the (system) stellar IMF, motivated by the remarkable observational resemblance between these two distributions. Note, however, that the bound cores identified in the simulations correspond to a time average, once each core becomes first bound, so that
identification with observed cores at a snapshot in time must be done with caution (see S09). Our results show that the various distributions of bound cores and sinks obtained in the simulations of cloud fragmentation are consistent with the sinks being formed preferentially from their parent prestellar core mass, {\it with some statistical dispersion} due e.g. to fluctuations in the core intrinsic properties (shape, density profile, velocity dispersion,...) and thus with the IMF being relatively tightly correlated to the CMF. Comparison with the results from Smith et al. (2009) allows us to quantify the characteristic standard deviation of this correlation, $\sigma\simeq \mcore/3$. Because of this dispersion,
we show that a core mass domain can contribute to different sink mass domains, with different statistical weights,
as obtained in the numerical simulations, although the final sink mass is built dominantly from its parent core mass reservoir.
The conclusion of this study is that the observed or numerically obtained similarity between the CMF and the IMF can be explained by  the IMF being determined predominantly by the CMF, within some
statistical fluctuations. This suggests that, at least within the mass domain explored in the present study, the newborn star accretes preferentially from the
mass reservoir in its immediate vicinity - i.e. its parent core - before accretion stops or decreases substantially, and that wider environmental factors or competitive accretion, although
probably partly contributing to the dispersion, are unlikely to be dominant factors to explain the observed properties. This also suggests that the dispersion during the first free-fall time or so in the simulations stems principally from {\it local} variations of the mass reservoir while the impact from wider environments becomes significant only at later times, as indeed suggested by S09. An argument raised by S09 is that long-term ($t\gg t_{ff}$) accretion can eventually modify the IMF and wipe out the initial (CMF) conditions. However, as seen in Fig.\ref{dist2}, even if accretion is pursued over several dynamical times, the CMF-IMF correlation is still of the order of about half a core mass, and thus is by no means completely wiped out. Moreover, such long lasting significant accretion seems unlikely (see e.g. Offner et al. (2009), Fig. 8). Indeed, there is observational evidence that accretion decreases significantly after the class-0 phase (about 1 to 2 $t_{dyn}$); furthermore, recent observations suggest  that the central protostar builts up essentially all its mass during a few episodes of violent accretion before this latter decreases substantially,
with about half the mass of a 0.5 $\msol$ dense prestellar core being accreted during less than 10\% of the Class I lifetime, i.e a fraction of a free-fall time (Evans et al. 2009). A scenario supported by numerical simulations (Vorobyov \& Basu 2006). Therefore, it seems unlikely that significant accretion lasts long enough for the
strong correlation between the (initial) CMF and the (final) IMF to be completely washed out.
This issue needs to be explored with dedicated numerical simulations including large- and
small-scale radiative and magnetic feedback processes.
In the same vein, the collision timescale between prestellar cores in dense star forming clumps appears to be significantly longer than the core lifetimes, suggesting that the cores should evolve individually to form  a small number of stars rather than competing for gas accretion, leading to a natural mapping of the CMF onto the IMF (Andr\'e et al. 2007, 2009; Evans et al. 2008).

According to the present analysis, the final stellar IMF can be determined reasonably accurately, on a {\it statistical} basis, i.e. not for individual objects, from the initial core mass distribution in the cloud, i.e. the CMF, with some unavoidable scatter, leading naturally to two similar mass distributions (see Fig. 3). Exploring a larger dynamical mass range with simulations, or conducting such a statistical analysis from {\it observed} CMF/IMF, a task possibly in reach with the HERSCHEL mission, would certainly help assessing this result. Such a correlation between the CMF and the IMF argues in favor of the IMF being essentially determined by the general properties of the parent cloud (mean temperature and density, large-scale velocity dispersion, with scaling properties following the Larson relations), as recently theorized by Hennebelle and Chabrier (2008,2009), and being only weakly affected by the various environmental factors beyond the parent core mass reservoir. Accordingly, the final IMF of a star forming region can be determined reasonably accurately, provided some average uniform efficiency factor, from the initial CMF obtained from mm- or sub-mm surveys. Interestingly enough, the present calculations also show that the aforementioned statistical variations among the core properties yield a final IMF extending further down in the low-mass domain than the parent CMF, providing a natural explanation for the fact that this latter appears to underestimate the number of "pre brown dwarf" cores compared with the observationally-derived brown dwarf IMF. These conclusions will have to be confronted to the wealth of data expected from the HERSCHEL mission, as nailing down this issue bears major consequences to understand the fundamental origin of star formation.

\acknowledgments

The authors are grateful to Rowan Smith, Ian Bonnell and Paul Clark for stimulating discussions and for sharing their data. We are also thankful to Philippe Andr\'e for fruitful conversations and insightful comments and to the anonymous referee for helping us improving the manuscript. We acknowledge funding from the European Research Council under the European Community 7th Framework Programme (FP7/2007-2013 Grant Agreement no. 247060.



\newpage

\end{document}